\documentclass[twoside]{article}

\usepackage{SIunits}

\usepackage{lipsum} 

\usepackage[sc]{mathpazo} 
\usepackage[T1]{fontenc} 
\usepackage{microtype} 

\usepackage[hmarginratio=1:1,top=32mm,columnsep=20pt]{geometry} 
\usepackage{multicol} 
\usepackage[hang, small,labelfont=bf,up,textfont=it,up]{caption} 
\usepackage{booktabs} 
\usepackage{float} 
\usepackage{hyperref} 

\usepackage{lettrine} 
\usepackage{paralist} 

\usepackage{abstract} 

\usepackage{titlesec} 
\renewcommand\thesection{\Roman{section}} 
\renewcommand\thesubsection{\Roman{subsection}} 
\titleformat{\section}[block]{\large\scshape\centering}{\thesection.}{1em}{} 
\titleformat{\subsection}[block]{\large}{\thesubsection.}{1em}{} 

\usepackage{fancyhdr} 
\usepackage[round, numbers,sort&compress]{natbib}
\usepackage{graphicx}

\title{\vspace{-10mm}\fontsize{22pt}{10pt}\selectfont\textbf{Analysis of whole mitogenomes from ancient samples}} 

\author{
\large
\textsc{Gloria G. Fortes \& Johanna L.A. Paijmans}\\[2mm] 
\normalsize Institute for Biochemistry and Biology, University of Potsdam, Potsdam, Germany \\ 
\small Corresponding Author: \href{mailto:paijmans.jla@gmail.com}{paijmans.jla@gmail.com} 
\vspace{-5mm}
}
\date{}

\begin{document}

\maketitle 

\thispagestyle{fancy} 


\begin{abstract}

\noindent Ancient mitochondrial DNA has been used in a wide variety of palaeontological and archaeological studies, ranging from population dynamics of extinct species to patterns of domestication. Most of these studies have traditionally been based on the analysis of short fragments from the mitochondrial control region, analysed using PCR coupled with Sanger sequencing. With the introduction of high-throughput sequencing, as well as new enrichment technologies, the recovery of full mitochondrial genomes (mitogenomes) from ancient specimens has become significantly less complicated. Here we present a protocol to build ancient extracts into Illumina high-throughput sequencing libraries, and subsequent Agilent array-based capture to enrich for the desired mitogenome. Both are based on previously published protocols, with the introduction of several improvements aimed to increase the recovery of short DNA fragments, while keeping the cost and effort requirements low. This protocol was designed for enrichment of mitochondrial DNA in ancient or degraded samples. However, the protocols can be easily adapted for using for building libraries for shotgun-sequencing of whole genomes, or enrichment of other genomic regions. 

\end{abstract}



\section{Introduction}

\lettrine[nindent=0em,lines=2]{T}he main characteristics of ancient DNA molecules are their low abundance in biological remains and the presence of molecular damage accumulated over time. Most of these damages are nicks caused by hydrolysis, which results in the progressive fragmentation of the DNA molecule into increasingly shorter fragments. The analysis of ancient samples is therefore particularly challenging, and for many years the principal focus of ancient DNA studies has been on mitochondrial DNA (mtDNA), as it is expected to be better preserved than nuclear DNA because of its higher copy number in the living cell. In such studies, short mtDNA fragments are typically targeted with specifically designed primer pairs and amplified for Sanger sequencing. However, there are a number of problems associated with this approach. For example, mismatches between the primers and target sequence can cause the reaction to work suboptimal (or even not at all). This is particularly problematic in the analysis of extinct species, or other species for which there is no reference available for primer design. Additionally, as most ancient DNA molecules are below 40 bp due to hydrolytic damage, successful amplification is challenging at best \citep{paabo_genetic_2004}. Finally, as modern contaminant DNA is generally longer, this approach is very sensitive to contamination, demanding replications of the experiments to validate their results \citep{gilbert_assessing_2005}. 

With the introduction of Next Generation Sequencing (NGS), the analysis of ancient DNA has undergone a tremendous advance. NGS involves the preparation of genomic libraries from the entire DNA extract by ligating adapters to each end. This means that species-specific primers are no longer necessary, effectively allowing every single fragment present in the extraction to be sequenced. Using such a shotgun-sequencing approach, the entire genomes from ancient modern humans \citep{rasmussen_ancient_2010,rasmussen_aboriginal_2011,keller_new_2012}, Neanderthals \cite{green_draft_2010}, Denisovans \cite{reich_genetic_2010,meyer_high-coverage_2012} and horse \cite{orlando_true_2011} have been sequenced. However, while whole ancient genomes are certainly possible these days, we should not disregard the power of mitogenomes. Mitogenomes represent an interesting and unique marker, particularly to address population genetics, and has been used in many ancient DNA studies \cite{paijmans_mitogenomic_2013}. For example, mitogenomes are still the marker of choice for samples with very poor preservation \cite{dabney_complete_2013,meyer_mitochondrial_2013}, or when comparing ancient samples to their known mitochondrial diversity \cite{zhang_morphological_2013,thalmann_complete_2013}.

As described above, the amount of DNA extracted from ancient samples is usually limited, with a low ratio of endogenous versus environmental and contaminant DNA. In general, the endogenous content is less than 1\% \cite[e.g.][]{orlando_true_2011} -- only in very rare cases this amount exceeds 50\%: under favourable preservation conditions \cite[e.g.][]{poinar_metagenomics_2006} or for particular sample types \cite[e.g.][]{gilbert_whole-genome_2007}. This means that when an ancient DNA library is directly sequenced, substantial sequencing is needed for each sample to recover sufficient coverage for the genomic region of interest -- an approach which is expensive and labour-intensive. 

To circumvent these problems, there are a number of methods available to enrich sequencing libraries for the desired DNA (\cite[also referred to as hybridization capture or just capture; for reviews see ][]{summerer_enabling_2009,mamanova_target-enrichment_2010}. This approach works by creating bait molecules with high sequence similarity to the target, which then hybridize to the desired DNA. Bait and target are then immobilized so that non-hybridized fragments can be washed away. The resulting library should then contain a much higher ratio of target versus contaminant DNA, so that subsequent sequencing yields much more usable data. Additionally, contrary to traditional PCR, hybridization capture favours smaller fragments \cite{hodges_genome-wide_2007} -- effectively introducing a bias against any modern contaminants, as these fragments are generally expected to be longer than ancient DNA. Many of the available capture protocols have already been applied to archaeological specimens, demonstrating the tremendous advantages of this approach \cite[e.g. ][]{zhang_morphological_2013,fu_dna_2013,avila-arcos_application_2011,burbano_targeted_2010,bos_draft_2011}. 

Here we present a protocol to build ancient DNA extracts into Illumina NGS libraries, suitable for shotgun sequencing of whole genomes, or enrichment of a particular target sequence. We subsequently describe a protocol to perform Agilent array-based capture to enrich for the complete mitochondrial sequence. Both are based on previously published protocols \cite[respectively]{meyer_illumina_2010,hodges_hybrid_2009}, with the introduction of several improvements aimed to: 1. increase the recovery of short DNA fragments, which is characteristically the majority of ancient DNA templates; and 2. optimize the pooling of multiple samples, in order to reduce the cost and time of laboratory procedures in ancient population studies. The protocols described below are performed on extracts generated by the silica-based spin column extraction method by Rohland et al. \cite{rohland_rapid_2010} -- however, the library preparation and capture is compatible with any extraction protocol that best suits the sample. 


\subsection{Illumina sequencing library construction}

The preparation of NGS libraries from ancient (or other degraded) DNA extracts requires protocols optimized for this particular purpose \cite{briggs_preparation_2012,bowman_multiplexed_2013,gansauge_single-stranded_2013}. In general, NGS library preparation methods require a number of consecutive enzyme reactions, and therefore include purification steps in between, either by solid phase reversible immobilization (SPRI) beads or silica-based filter columns. These purification steps may cause significant loss of template -- particularly the shortest fragments, which is particularly detrimental in the case of ancient samples. A recently published library preparation method has proven very effective for ancient samples, mainly because of the elimination of purification steps \cite{gansauge_single-stranded_2013}. However, the single-stranded library preparation method is particularly costly and time consuming, making the double-stranded protocols generally still the method of choice for projects with larger number of samples. 

The library preparation protocol described here is based on the original protocol for home-made Illumina libraries presented by Meyer and Kircher \cite{meyer_illumina_2010}, with three main adaptations for increased sensitivity. First, we have reduced the number of purification steps from three down to only one, by replacing the column-based purification by heat inactivation of the enzymes (Fig. 1). This modification helps to avoid the loss of the shortest DNA fragments, particularly after the first enzyme reaction (blunt-end repair), at which stage DNA is not yet ligated to the adapters and the risk of losing the smallest fragments is highest. Another adaptation in our library preparation is the use of a second index on the 5' end of the ancient DNA template. The importance of such double-indexing has been discussed previously \cite{craig_identification_2008,kircher_double_2011}. Double-indexing serves as an extra identification of the sequence, and thus allows for pooling of more samples on a single lane. In addition to this, since the indices are located at both ends of the template, double-indexing also provides a means of filtering chimeric reads (or jumping-PCR) from the dataset, and thus increases the confidence with which a particular read can be assigned to the corresponding library. The indexing method that we utilise does not require an additional sequence read; the P5 index is retrieved as part of the regular read 1 (Fig. 1, Note 1). Finally, we introduced parallel PCR's to maintain more complexity of the library during amplification. As the amount of endogenous template molecules in ancient samples is generally low, the amplification of the library can introduce biases (i.e. preferential amplification of certain fragments). We reduce this loss of complexity by amplifying every sample in a number of parallel PCR reactions (usually four), each containing a unique subset of the original library as starting templates.

\begin{figure}[h!]
  \centering
    \includegraphics[width=0.5\textwidth]{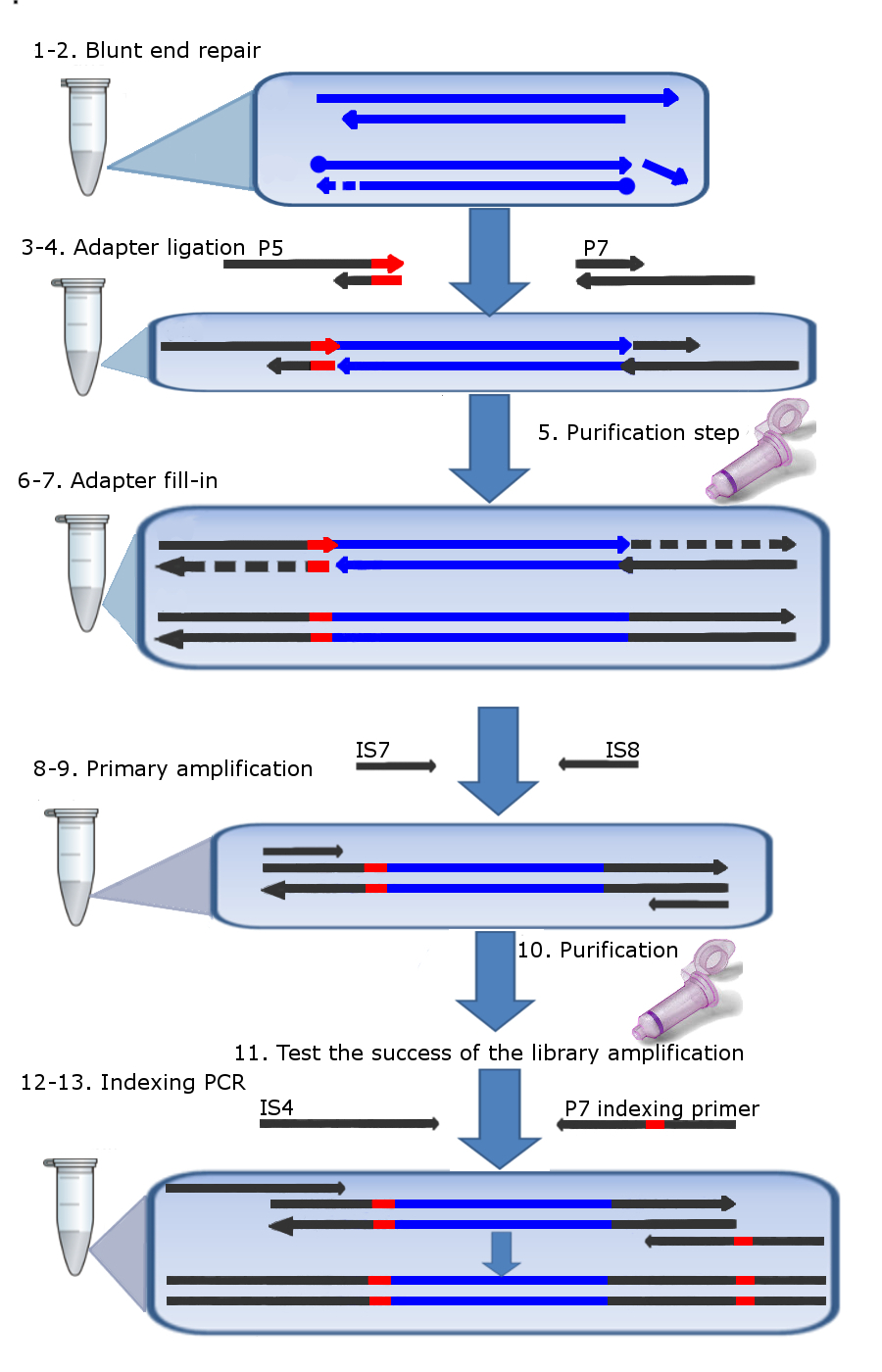}
    \caption{Library preparation procedure and amplification. 1-2. The first step is the blunt-end repair, which removes any overhangs resulting from hydrolytic damage. The T4 DNA polymerase fills in or removes 5' and 3' overhangs, respectively, while T4 PNK polymerase phospohorylates the 5' ends (round edges). 3-4. The Illumina-specific adapters (P5 and P7) are ligated to the template molecule by the T4 DNA ligase. As the P5 adapter includes an index sequence (position indicated in red), the library is already indexed from this step onwards. 5. The adapters that are not incorporated into the library, will be discarded during the purification using silica-based columns (MinElute PCR Purification columns). 6-7. The fill-in reaction uses a Bst polymerase with strand displacement activity that fills in the nicks and completes the double strand sequence of the adapter. 8-9. A primary PCR with primers IS7 and IS8 is used to amplify the library in a quantity that suits the requirements of the capture procedure. 10. The excess of primers will be discarded during purification using MinElute PCR purification columns. 11. There are several options to test the success of the primary amplification; either by electrophoresis on a 2.5\% Agarose gel, or Agilent Bioanalyser/Tapestation. 12-13. During the indexing PCR, 5'-tailed primers (IS4 and P7 index primer) are used to reconstruct the full sequence of the Illumina-specific adapters. The red label in the P7 indexing primer represents the position of the P7-index. When more than one sample is built into library, each of the described steps has to be carried out separately for each sample. Only after the indexing PCR, the libraries are ready to be pooled. After pooling the libraries, amplification can be performed using the IS5 and IS6 oligonucleotides (oligos). All oligo sequences can be found in Table 1.}
\end{figure}


\subsection{DNA hybridization capture}

The enrichment method described in this protocol is a solid-state capture utilizing high-density tiling DNA microarrays to immobilize target regions \cite[after][]{hodges_genome-wide_2007,hodges_hybrid_2009}. Microarray capture has been successfully applied to  archaeological specimens \cite{avila-arcos_application_2011,burbano_targeted_2010,burbano_analysis_2012} and has some advantages over in-solution capture. Solid-state capture can be particularly interesting when the target region is large (i.e. >1 Gb), since in such cases buying or creating the baits is rapidly becoming expensive and labour-intensive. In addition, contrary to many in-solution capture protocols, array capture avoids unequal representation of baits (e.g. repetitive regions), which can lead to uneven or patchy coverage of the target regions \cite{hodges_hybrid_2009}. 
The protocol described here is based on the protocol by Hodges et al. \cite{hodges_hybrid_2009}, with four adaptations to facilitate efficiency and sensitivity (Fig. 2). Firstly, we have eliminated the species-specific COT I DNA in the hybridization mixture. In the original protocol COT I is added to prevent over-representation of repetitive elements, which can be introduced by spurious hybridization of template molecules to the overhang of captured molecules instead of baits \cite{hodges_hybrid_2009}. However, we have found that exuding COT I does not have an adverse effect when capturing ancient DNA, possibly because ancient DNA fragments rarely exceed the bait length of 60 bp. Secondly, we exclude the concentration of the eluate using a SpeedVac, to prevent any potential loss of template during this step. Third, after elution of the enriched library from the array, we have introduced additional parallel PCR reactions -- both to account for the increased elution volume due to the elimination of the SpeedVac step, as well as to maintain complexity of the enriched libraries during amplifications as discussed above. Finally, it has been found that performing a second capture on the already enriched material will increase the enrichment rate \cite{li_capturing_2013} -- therefore we have introduced a second capture round in our protocol, identical to the first enrichment. While this step does increase the time and money required for the enrichment, this does not outweigh the increased yield of usable sequence data. We have also found that, provided that the array is stored dry and not too long, it can be re-used a second time for capture. The protocol described here has been applied successfully to capture mitogenomes as well as nuclear DNA. 

\begin{figure}[h!]
  \centering
    \includegraphics[width=0.5\textwidth]{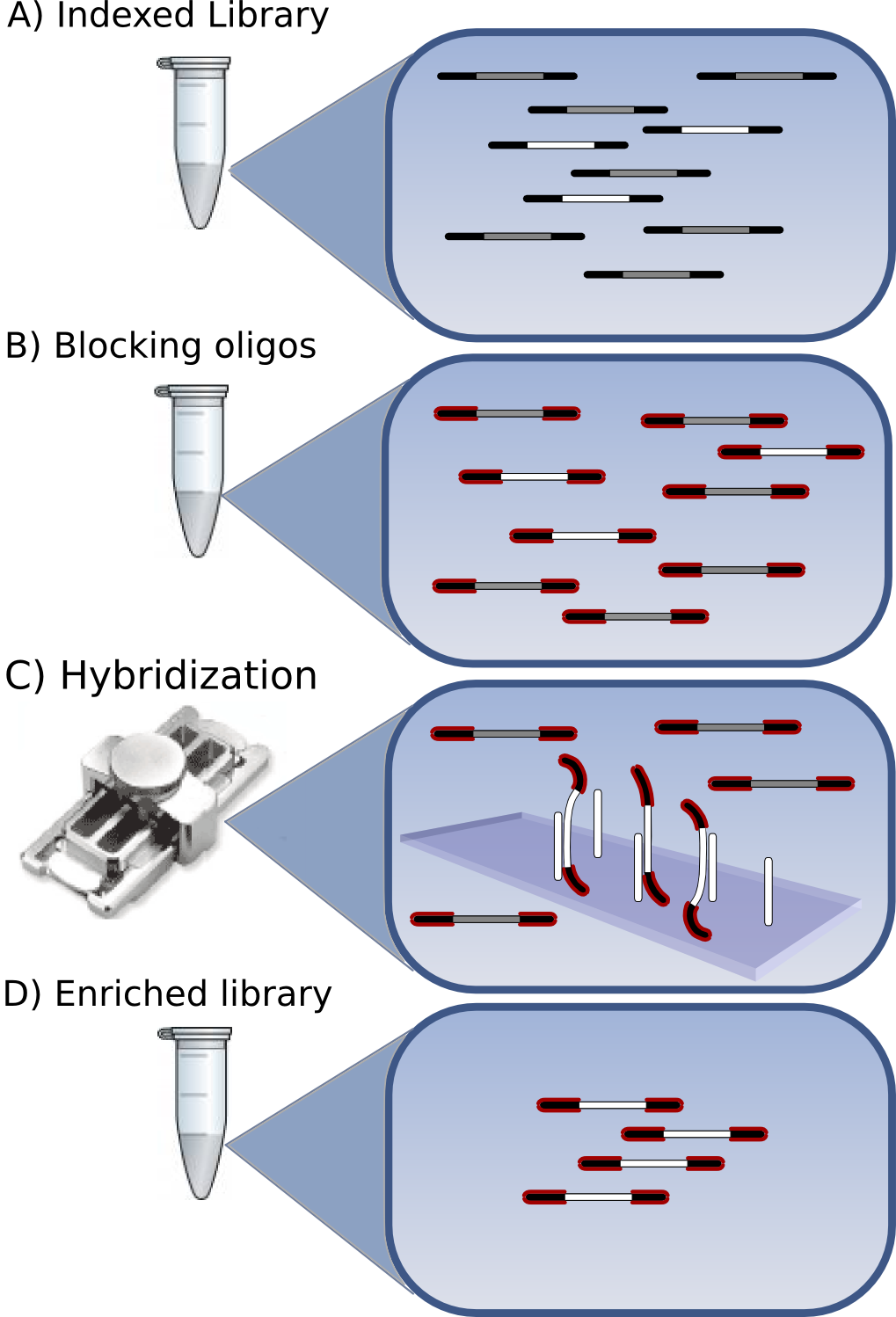}
    \caption{Solid-state hybridization capture using a custom-designed Agilent microarray. The capture is performed either on pooled and indexed libraries, or on primary, pre-indexing libraries (A). Blocking oligos that correspond to the adapter sequences are added to prevent these from interfering during the hybridization capture (B). Then, the mixture is loaded on the array and hybridized for 60-65 h (C). The result is a library enriched for the target region (in this case, mitogenomes) (D).}
\end{figure}


\section{Materials}

\begin{enumerate}
\item  Oligonucleotides (oligos) as adapted from Meyer and Kircher, 2010 (Note 1--5):

\begin{table}[H]
\footnotesize
\caption{Sequences for all oligonucleotides required for hybridization capture \cite[originally described in ][]{meyer_illumina_2010}}. 
\centering
\begin{tabular}{lll}
\toprule
Name & Sequence & Notes \\
\midrule
P5\_ind\_F\_xxx & \scalebox{.8}[1.0]{A*C*A*C*TCT TTC CCT ACA CGA CGC TCT TCC GAT CT xxxx*x*x*x*x} & "x" indicates unique index sequence (Note 19) \\ 
 & & * phosphathioate oligonucleotide (PTO) bond \\
P5\_ind\_R\_xxx & \scalebox{.8}[1.0]{x*x*x*x*xxxx AGA*T*C*G*G} & "x" indicates unique index, * indicates PTO bond \\
P7\_adapter\_F & \scalebox{.8}[1.0]{G*T*G*A*CTG GAG TTC AGA CGT GTG CTC TTC CG*A*T*C*T} & * indicates PTO bond \\
P7\_adapter\_R & \scalebox{.8}[1.0]{A*G*A*T*CGGAA*G*A*G*C} & * indicates PTO bond \\
IS4 & \scalebox{.8}[1.0]{ - AAT GAT ACG GCG ACC ACC GAG ATC TAC ACT -} & \\
 & \scalebox{.8}[1.0]{\hspace{1cm}CTT TCC CTA CAC GAC GCT CTT} & \\
IS5 & \scalebox{.8}[1.0]{AAT GAT ACG GCG ACC ACC GA} & \\
IS6 & \scalebox{.8}[1.0]{CAA GCA GAA GAC GGC ATA CGA} & \\
IS7 & \scalebox{.8}[1.0]{ACA CTC TTT CCC TAC ACG AC} & \\
IS8 & \scalebox{.8}[1.0]{GTG ACT GGA GTT CAG ACG TGT} & \\
Indexing\_primer\_xxx & \scalebox{.8}[1.0]{CAA GCA GAA GAC GGC ATA CGA GAT xxxxxxxx -} & "x" indicates unique index sequence \\
 & \scalebox{.8}[1.0]{\hspace{1cm} - GTG ACT GGA GTT CAG ACG TGT} & \\
BO P5\_trun\_F & \scalebox{.8}[1.0]{ACA CTC TTT CCC TAC ACG ACG CTC TTC CGA TCT-Pho} & Pho indicates a 3'-end phosphate \\
BO P5\_trun\_R & \scalebox{.8}[1.0]{AGA TCG GAA GAG CGT CGT GTA GGG AAA GAG TG-Pho} & 3'-end phosphate \\
BO P5\_ext\_F & \scalebox{.8}[1.0]{ATC TCG TAT GCC GTC TTC TGC TTG-Pho} & 3'-end phosphate \\
BO P5\_ext\_R & \scalebox{.8}[1.0]{CAA GCA GAA GAC GGC ATA CGA GAT-Pho} & 3'-end phosphate \\
BO P7\_trun\_F & \scalebox{.8}[1.0]{AGA TCG GAA GAG CAC ACG TCT GAA CTC CAG TCA C-Pho} & 3'-end phosphate \\
BO P7\_trun\_R & \scalebox{.8}[1.0]{GTG ACT GGA GTT CAG ACG TGT GCT CTT CCG ATC T-Pho} & 3'-end phosphate \\
BO P7\_ext\_F & \scalebox{.8}[1.0]{ATC TCG TAT GCC GTC TTC TGC TTG-Pho} & 3'-end phosphate \\
BO P7\_ext\_R & \scalebox{.8}[1.0]{CAA GCA GAA GAC GGC ATA CGA GAT-Pho} & 3'-end phosphate \\
\end{tabular}
\end{table}

\item Reagents

\begin{enumerate}
 \item[1.] Oligo hybridization buffer (10x): 500 mM NaCl, 10 mM Tris-HCl (pH 8.0), 1 mM EDTA (pH 8.0)
 \item[2.] Restriction buffer (10x Buffer Tango, e.g. Fermentas)
 \item[3.] T4 DNA ligase, with corresponding buffer and PEG-4000 (5 U/\micro L, e.g. Fermentas)
 \item[4.] T4 DNA polymerase (5 U /\micro L e.g. Fermentas)
 \item[5.] T4 polynucleotide kinase (10 U/\micro L, e.g. Fermentas)
 \item[6.] ATP (100 mM)
 \item[7.] Bst DNA polymerase, large fragment, supplied with corresponding Thermopol Reaction Buffer (e.g. New England BioLabs)
 \item[8.] AmpliTaq Gold (Applied Biosystems, supplied with 10x AmpliTaq buffer and 25 mM MgCl2)
 \item[9.] dNTP's (e.g. Invitrogen)
 \item[10.] Bovine Serium Albumin (BSA, e.g. Promega)
 \item[11.] Agilent aCGH/ Chip-on-Chip Wash Buffer 1 (Agilent, cat no. 5188-5221)
 \item[12.] Agilent aCGH/ Chip-on-Chip Wash Buffer 2 (Agilent, cat no. 5188-5222)
 \item[13.] Oligo aCGH \& ChIP-on-Chip Hybridization Kit (Agilent, cat no. 5188-5220), contains 2X Hi-RPM Hybridization Buffer and 10X Oligo aCGH/ChIP-on-Chip Blocking Agent. The Blocking agent needs to be resuspended by adding 1,350 \micro L nuclease-free water. Incubate at room temperature for 60 min and mix gently.
 \item[14.] Qiagen MinElute PCR purification columns (Qiagen, cat no. 28004), contains Qiagen Minelute purification spin columns, PB binding buffer, PE washing buffer and EB elution buffer, as well as Loading dye and pH indicator (not used). Before use, add ethanol (96-100\%) to PE buffer (see bottle label for volume).

\end{enumerate}

\item Equipment

\begin{enumerate}
 \item[1.] Lab suitable for ancient DNA work: i.e. physically isolated from labs where work on modern DNA is performed, sterile working environment and consumables, appropriate protective clothing.
  \item[2.] Thermal cycler in cleanlab and modern lab (e.g. Veriti Thermal Cycler, Life Technologies).
  \item[3.] Microcentrifuge in cleanlab and modern lab (e.g. Labnet Spectrafuge 24D).
  \item[4.] Microarray hybridization oven, temperature range up to 95\degree C (section 9 and 18; e.g. Affymetrix Model 777 Hybridization Oven) (Note 6).
  \item[5.] Incubator oven (37\degree C; e.g. Labnet Mini Incubator with Mini Lab Roller).
  \item[6.] SureHyb DNA Microarray Chamber (step 4-19) (Agilent).
  \item[7.] Microarray gasket slides (Agilent).
  \item[8.] Slide spinner or centrifuge fitted with microplate adapters.
  \item[9.] Slide rack.
  \item[10.] 3 slide-staining dishes.
  \item[11.] 1 big dish (big enough to envelop one of the slide staining dishes with sufficient room to spare).
  \item[12.] Heating blocks (95\degree C and 37\degree C).
  \item[13.] Magnetic hot plate and stirrers.
  \item[14.] 30G needles and syringes.
\end{enumerate}

\item Preparation of indexed adapter mixes

\begin{enumerate}
\item[1.] Prepare the P5 and P7 adapter mixes as follows:
	\begin{enumerate}
 	\item[1.] P5 indexed adapter mix ('xxx' indicates index identifier):
		\begin{enumerate}
    	\item[1.] 1 \micro L of oligo P5\_index\_F\_xxx (500 \micro M)
 		\item[2.] 1 \micro L of oligo P5\_index\_R\_xxx (500 \micro M)
 		\item[3.] 1.25 \micro L of Oligo hybridization buffer (10x)
 		\item[4.] 9.25 \micro L of H2O
 		\end{enumerate}
	\item[2.] P7 adapter mix
		\begin{enumerate}
 		\item[1.] 1 \micro L of oligo P7\_adapter\_F (500 \micro M)
 		\item[2.] 1 \micro L of oligo P7\_adapter\_R (500 \micro M)
 		\item[3.] 1.25 \micro L of Oligo hybridization buffer (10x)
 		\item[4.] 9.25 \micro L of H2O
	\end{enumerate}
	\item[2.] Incubate the two mixtures for 10 sec at 95\degree C, followed by a ramp from 95\degree C to 12\degree C at a rate of 0.1\degree C/sec.
	\item[3.] Pool the two mixtures (25 \micro L total volume), resulting in a ready-to-use indexed adapter mix containing 20 \micro M of each adapter.
	\item[4.] Store adapter mixes at -20\degree C.
	\end{enumerate}
\end{enumerate}
\end{enumerate}

 
\section{Methods}

\begin{enumerate}
{\bf \large \item Array design (Note 7)}
	\begin{enumerate}
	\item[1.]  Retrieve the mitogenome for the desired species (or close relative) from GenBank (http://www.ncbi.nlm.nih.gov/nucleotide/) -- if more than one sequence is available, it is advisable to include them all in the array-design to maintain the range of known sequence diversity of the species of interest. 
	\item[2.] Use RepeatMasker (http://www.repeatmasker.org/; A.F.A. Smit et al., unpublished data) to mask repeats. In some cases, the tRNA's will be masked as well, if desired these can be unmasked and included in the baits.
	\item[3.] Create baits of 60 bp length with 1 or 2 bp tiling space, and use these to design the Agilent SureSelect DNA capture array (either 244k or 1M features). 
	\end{enumerate}


{\bf \large \item Library preparation}

\begin{enumerate}
	\item[1.] Set up a master mix for blunt-end repair. Include per reaction (Note 8):
    	\begin{enumerate}
		\item[1.] 3.56 \micro L of H2O
		\item[2.] 3.5 \micro L of Buffer Tango (10x)
		\item[3.] 0.14 \micro L dNTP's (25 mM each)
		\item[4.] 0.25 \micro L of ATP (100 mM)
		\item[5.] 1.75 \micro L of T4 Polynucleotide kinase (10 U/\micro L)
		\item[6.] 0.7 \micro L of T4 Polymerase (5 U/\micro L)
        \end{enumerate}
	\item[2.] Add 10 \micro L of blunt-end repair master mix to 25\micro L of ancient DNA extract. Mix gently by pipetting up and down, and incubate in a thermal cycler at 25\degree C for 20 min, followed by heat inactivation step at 72\degree C for 20 min, then hold at 12\degree C.
	\item[3.] Set up a master mix for adapter ligation, with per reaction:
		\begin{enumerate}
        \item[1.] 2.5 \micro L of H2O
		\item[2.] 5 \micro L of T4 ligase buffer (10x)
		\item[3.] 5 \micro L of PEG-4000 (50\%)
		\item[4.] 1.25 \micro L of T4 Ligase (5 U/\micro L)
        \end{enumerate}
	\item[4.] Add 1.25 \micro L of a unique indexed adapter mix to each 35 \micro L sample and mix carefully but thoroughly. Then add 13.75 \micro L of adapter ligation master mix to each sample (Note 9). Mix gently and incubate in a thermal cycler at 22\degree C for 30 min, then hold at 12\degree C.
	\item[5.] Purify the reaction using a MinElute PCR purification column. Add per reaction 100 \micro l Qiagen PB buffer. Centrifuge for one minute and discard flow-through. Add 750 \micro l Qiagen PE washing buffer and centrifuge again at the previously given speed and time. Discard flow-through and repeat centrifugation. Place the Qiagen MinElute in a new tube, and add 10\micro L Qiagen EB buffer. Incubate for 5 minutes and centrifuge at the same speed and time. Add another 10\micro l of EB buffer and incubate for 5 minutes, and centrifuge. All centrifugation steps are carried out at 17,900 x g (13,000 rpm) in a tabletop microcentrifuge at room temperature. 
	\item[6.] Set up a master mix for adapter fill-in, with per reaction:
		\begin{enumerate}
        \item[1.] 14.1 \micro L of H2O
		\item[2.] 4 \micro L of Thermopol buffer (10x)
		\item[3.] 0.4 \micro L of dNTP's (25 mM each)
		\item[4.] 1.5 \micro L of Bst Polymerase (8 U/\micro L)
        \end{enumerate}
 7. Add 20 \micro L of adapter fill-in master mix to each 20 \micro L sample. Mix gently and incubate in a thermal cycler at 37\degree C for 20 min, followed by heat inactivation step at 80\degree C for 20 min, then hold at 12\degree C
	\item[8.] For the primary amplification of the libraries, set up a primary amplification master mix as follows, with per parallel reaction:
		\begin{enumerate}
        \item[1.] 7.8 \micro L of H2O
		\item[2.] 2 \micro L of AmpliTaq Gold buffer (10x)
		\item[3.] 1.6 \micro L of MgCl2 (25 mM)
		\item[4.] 1.5 \micro L of primer IS7 (10 \micro M)
		\item[5.] 1.5 \micro L of primer IS8 (10 \micro M)
		\item[6.] 0.2 \micro L of BSA (10 mg/mL)
		\item[7.] 0.2 \micro L of dNTP's (25 mM each)
		\item[8.] 0.2 \micro L AmpliTaq Gold Polymerase (5 U/\micro L) (Note 10)
        \end{enumerate}
	\item[9.] Add 15 \micro L of master mix to each 5 \micro L sample (Note 11). Mix gently and amplify according to the following temperature profile: initial denaturation 94\degree C for 10 min, followed by 15 cycles of 94\degree C for 30 sec, 60\degree C for 45 sec, and 72\degree C for 45 sec, followed by a final extension at 72\degree C for 5 min (Note 12).
	\item[10.] Purify the pooled parallel reactions using a MinElute PCR purification column, according to the manufacturers' instructions. Elute twice with 10 \micro L EB buffer, each with 5 min incubation time. 
	\item[11.] Verify the success of the primary amplification on a 2.5\% Agarose gel or Agilent Bioanalyser/Tapestation.
	\item[12.] Set up the master mix for the indexing PCR, with per parallel reaction:
		\begin{enumerate}
        \item[1.] 9.3 \micro L of H2O
		\item[2.] 2 \micro L of AmpliTaq Gold buffer (10x)
		\item[3.] 1.6 \micro L of MgCl2 (25 mM)
		\item[4.] 1.5 \micro L of primer IS4 (10 \micro M)
		\item[5.] 0.2 \micro L of BSA (10 mg/mL)
		\item[6.] 0.2 \micro L of dNTP's (25 mM each)
		\item[7.] 0.2 \micro L AmpliTaq Gold Polymerase (5 U/\micro L) (Note 10)
        \end{enumerate}
	\item[13.] Add 1.5 \micro L of a unique index primer (10 \micro M) to each 5 \micro L sample (Note 11) and 13.5 \micro L of master mix. Mix gently and amplify according to the following temperature profile: initial denaturation 94\degree C for 10 min, followed by 10 cycles of 94\degree C for 30 sec, 60\degree C for 45 sec, and 72\degree C for 45 sec, followed by a final extension at 72\degree C for 5 min (Note 12).
	\item[14.] Purify the pooled parallel reactions using a MinElute PCR purification column, as described in section 3.5.
	\item[15.] Verify the success of the primary amplification on a 2.5\% Agarose gel or Agilent Bioanalyser/Tapestation.
\end{enumerate}

{\bf \large \item  Capture}

\begin{enumerate}

\item[1.] Pool the libraries in equimolar amounts (Note 13,14) and bring the volume up to 168 \micro L with H2O.
\item[2.] Prepare hybridization mix as follows (maintain the order in which they are added):
	\begin{enumerate}
    \item[1.] 168 \micro L of the pooled library
	\item[2.] 5 \micro L of oligo BO\_P5\_trun\_F (200 \micro M)
	\item[3.] 5 \micro L of oligo BO\_P5\_trun\_R (200 \micro M)
	\item[4.] 5 \micro L of oligo BO\_P5\_ext\_F (200 \micro M)
	\item[5.] 5 \micro L of oligo BO\_P5\_ext\_R (200 \micro M)
	\item[6.] 5 \micro L of oligo BO\_P7\_trun\_F (200 \micro M)
	\item[7.] 5 \micro L of oligo BO\_P7\_trun\_R (200 \micro M)
	\item[8.] 5 \micro L of oligo BO\_P7\_ext\_F (200 \micro M)
	\item[9.] 5 \micro L of oligo BO\_P7\_ext\_R (200 \micro M)
	\item[10.] 52 \micro L of Agilent Blocking agent (10x)
	\item[11.] 260 \micro L of Agilent Hybridization buffer (2x)
    \end{enumerate}
\item[3.] Incubate the mixture at 95\degree C for 3 min, followed by 37\degree C for 30 min (Note 15)
\item[4.] Disassemble the hybridization chamber and place a clean, dry gasket in the chamber.
\item[5.] Load 490 \micro L of the mixture on the gasket (Note 16)
\item[6.] Slowly load the array down on the gasket (numbered barcode facing up, Agilent-labeled barcode facing down (Reminder: Agilent = Active)
\item[7.] Reassemble the chamber, screw tight to secure.
\item[8.] Make sure there are no stationary bubbles by rotating the chamber. If there are stationary bubbles, tap the chamber on a solid surface to dislodge them.
\item[9.] Hybridize for 60-65 h at 65\degree C under rotation at 12 rpm.
\item[10.] After 48 h (i.e. the day before the hybridization is complete), preheat the materials necessary for the washing step with washing buffer 2 (see step l). Fill one bottle with H2O, and one bottle with Agilent washing buffer 2, and incubate overnight at 37\degree C together with the glassware.
\item[11.] Disassemble the hybridization chamber and move the whole gasket-array sandwich into a dish with washing buffer 1. Pry the slides apart (Note 17) and move the array to the slide holder and washing dish 1. Make sure the slide is not exposed to air for extended periods of time.
\item[12.] With gentle stirring, incubate at room temperature for 10 min.
\item[13.] Prepare the washing dish 2: put the small dish in the big dish. Fill the big dish with water and the small dish with Agilent washing buffer 2. This way, the washing buffer will maintain a constant temperature of 37\degree C.
\item[14.] Move the slide holder with the array into the dish containing washing buffer 2
\item[15.] With gentle stirring, incubate at 37\degree C for 5 min.
\item[16.] Spin the slide for 1 min at 600 rpm to dry.
\item[17.] Load 490 \micro L nuclease-free water on a new gasket, and reassemble the hybridization chamber. Again, make sure there are no stationary bubbles.
\item[18.] Incubate for 10 min at 95\degree C. If the temperature drops after opening the oven, start timer when oven reaches 90\degree C.
\item[19.] Use a needle and syringe to extract the eluate: first, slightly loosen the screw of the hybridization chamber. Then push the needle through the rubber seal on the non-labelled side of the slides. If the needle does not go through, loosen the screw of the chamber a bit more and try again. Elute the entire volume from the array before disassembling the chamber. 
\item[20.] After capture amplify the enriched library pool in 24 parallel reactions. Set up the amplification master mix as follows, with for each reaction (Note 18):
	\begin{enumerate}
	\item[1.] 5.6 \micro L of H2O
	\item[2.] 4 \micro L of AmpliTaq Gold buffer (10x)
	\item[3.] 3.2 \micro L of MgCl2 (25 mM)
	\item[4.] 3 \micro L of primer IS5 (10 \micro M)
	\item[5.] 3 \micro L of primer IS6 (10 \micro M)
	\item[6.] 0.4 \micro L of BSA (10 mg/mL)
	\item[7.] 0.4 \micro L of dNTP's (25 mM each)
	\item[8.] 0.4 \micro L AmpliTaq Gold Polymerase (5 U/\micro L)
    \end{enumerate}
\item[21.] Add 20 \micro L of master mix to each 20 \micro L sample. Mix gently and amplify according to the following temperature profile: initial denaturation 94\degree C for 10 min, followed by 20 cycles of 94\degree C for 30 sec, 60\degree C for 45 sec, and 72\degree C for 45 sec, followed by a final extension at 72\degree C for 5 min.
\item[22.] Purify the pooled parallel reactions using three MinElute PCR purification columns, as described in section 3.5. Elute every column twice with 10\micro L EB buffer, each with 5 min incubation time.
\item[23.] Verify the success of the primary amplification on a 2.5\% Agarose gel or Agilent Bioanalyser/Tapestation.
\item[24.] Repeat the capture (starting from step 3.1) (Note 19). 

\end{enumerate}
{\bf \large \item Notes}

\begin{enumerate}
	\item[1.] Interesting to note: because of the P5 index in-line at the beginning of read 1, the libraries can also be pooled and sequenced on a single-end sequence run and still be de-multiplexed afterwards. This is particularly interesting for ancient DNA, where paired-end sequencing in order to access longer sequences is not particularly useful.
	\item[2.] The blocking oligos mentioned here would also be compatible with the indexing method described in (32), as it has two independent blocking oligos for the P5 adapter.
	\item[3.] The indices are not actually blocked. However, because the indices are only 8 bp long, any spurious hybridization will be disrupted during the heated washing step.
	\item[4.] It is important to pay particular attention to the choice of P5 indices for samples that will be sequenced on a single lane: since the Illumina technology bases cluster calling on the first cycles of R1, the indices must have sufficient base diversity from each other to facilitate this process. When a low number of samples with limited diversity between the indices are sequenced on one lane, it is advisable to add PhiX library (a proprietary control library, available from Illumina) to the sample to increase the base diversity in the first cycles.
	\item[5.] When designing the index sequences, several considerations must be taken into account. The indices must be as different as possible, at least by three substitutions. The differences serve to avoid misidentification of the index sequence, either due to misincorporation of the nucleotides during library amplification or to sequencing errors. Illumina instruments only have two lasers, red for A/C and green for G/T; therefore for each position, the number of A and C bases should be well balanced with the number G and T bases for all indices used in the same experiment. For more information, see \cite{meyer_illumina_2010}. 
    \item[6.] For step 3.18; if no oven is available that rotates and goes up to 95\degree C, an alternative is to put the chamber on one side, incubate for 5 min, rotate to the other side and incubate for another 5 min.
	\item[7.] The array design described here is meant to enrich the library for whole mitogenomes: if other sequences are desired, one can simple follow the same procedure with the desired sequence instead of the mitogenome.
	\item[8.] We recommend counting one more reaction than the actual number of samples due to volume loss during pipetting
	\item[9.] To avoid the formation of adapter dimers, it is important to add the adapters to the sample, mix carefully by pipetting up and down, and only then add the master mix. If there are many adapter dimers in the final library, less adapter mix can be used. However, as this is blunt-ended ligation, adding too little of the adapter mix can cause templates to ligate to each other instead of the adapters, causing a loss of template (33).
	\item[10.] We have observed that the use of Accuprime Supermix I enzyme (Invitrogen) reduces the formation of adapter- and primer-dimers compared to AmpliTaq Gold. Like AmpliTaq Gold, Accuprime Supermix I includes a polymerase that can read over uracils, so it can also be used for the primary amplification of ancient DNA templates. However, it is recommended for amplicons < 200 bp, so its use should be avoided when inserts over 100 bp are expected.
	\item[11.] These amounts are based on four parallel reactions per sample. If a different number of parallel reactions are required, please adjust the amounts accordingly.
	\item[12.] If the library is not visible or weak, increase cycle numbers to 20. Alternatively, keep the number of cycles to 15 and reduce the elution volume to 10-12 \micro L of EB buffer. This will help to visualize the library, while keeping low the risk of clonality.
	\item[13.] Be aware when calculating the equimolar amounts of samples containing adapter dimers -- use the total concentration, not the concentration of individual bands. While theoretically the adapter dimers should not hybridize to the baits and be washed away, in practice a certain amount of dimers will still carry over to the post-capture library. When there is access to a Pippin prep (Target size selection platform by Sage Science, Beverly MA), this problem can be solved. We would not recommend using gel extraction for size selection of the library, as in our experience it does not work very well.
	\item[14.] Anywhere between 0.5 and 20 \micro g final amount of DNA should work for enrichment, but the more the better.
	\item[15.] Masking tape should be used to close the lid of the tube during the 95\degree C incubation; parafilm may not be sufficient to keep the lid closed.
	\item[16.] Distribute the mixture equally over the surface while keeping away from the rubber seal around the edges.
	\item[17.] If the slides are difficult to pry apart, try sticking a needle through the rubber seal to equalize the pressure.
	\item[18.] When performing the protocol for the first time or when the results are hard to predict, we advise amplifying eight parallel PCR reactions first, and then the remaining 16. By doing so, one still has template when optimization of the master mix, cycle numbers, etc. is necessary.
	\item[19.] During the final amplification after the second capture round, it may be wise to set up 8 parallel reactions first, purifying them in one MinElute column, and then verifying the success of the reaction. If the library is too bright or too weak, adjust cycle numbers. Then amplify the remaining 16 parallels and then purify using 2 MinElute columns.

\end{enumerate}
\end{enumerate} 


\newpage
\bibliographystyle{bibgen}
\bibliography{My_Library.bib}



\end{document}